\documentclass{pasj00}
\draft
\Received{2012/8/22}
\Accepted{2012/10/19}
\Published{2013/4/25}
\SetRunningHead{Astronomical Society of Japan}
{VLBI Imagings of Kilo-parsec Knot in 3C~380}

\begin{document}

\title{VLBI Imagings of Kilo-parsec Knot in 3C~380}
\author{Shoko Koyama\altaffilmark{1,2}, 
Motoki Kino\altaffilmark{3}, 
Hiroshi Nagai\altaffilmark{2}, 
Kazuhiro Hada\altaffilmark{4}, 
Seiji Kameno\altaffilmark{5},
and Hideyuki Kobayashi\altaffilmark{1,2}}

\altaffiltext{1}{Department of Astronomy, Graduate School of Science, The University of Tokyo, 7-3-1 Hongo, Bunkyo-ku, Tokyo 113-0033, Japan} 
\email{shoko.koyama@nao.ac.jp}
\altaffiltext{2}{National Astoronomical Observatory of Japan, 
2-21-1 Osawa, Mitaka, Tokyo 181-8588, Japan}
\altaffiltext{3}{The Institute of Space and Astronautical Science, Japan Aerospace Exploration Agency, 3-1-1 Yoshinodai, Chuou-ku, Sagamihara, Kanagawa 252-5210, Japan}
\altaffiltext{4}{INAF Istituto di Radioastronomia, via Gobetti 101, 40129 Bologna, Italy}
\altaffiltext{5}{Department of Physics, Faculty of Science, 
Kagoshima University, 1-21-35 Korimoto, Kagoshima 890-0065, Japan}


\maketitle

\begin{abstract}

We investigate observational properties of a
kilo-parsec scale knot in radio-loud quasar 3C~380 
by using two epoch archival data 
obtained by Very Long Baseline Interferometry (VLBI) at 5~GHz on 1998 July and 2001 April.
We succeed in obtaining
the highest spatial resolution image of the bright
knot K1 located at 732~milliarcseconds, or $\geqq$20~kpc de-projected,
downstream from the nucleus
three times better than previously obtained 
highest resolution image by Papageorgiou et al. (2006).
Our images reveal, with new clarity,
$``$inverted bow-shock" structure in K1
facing the nucleus and its morphology 
resembles a conical shock wave.
By comparing the two epoch images directly,
we explore the kinematics of K1 and 
obtain the upper limit of apparent velocity, $0.25~{\rm mas~yr^{-1}}$ or $9.8c$ of K1 for the first time.
The upper limit of apparent velocity is marginally smaller than 
superluminal motions seen in the core region.
Further new epoch VLBI observations
are necessary to measure the proper motion at K1.
\end{abstract}


\section{Introduction}

Thanks to recent progress in the field of radio interferometry,
the properties of knots at large scales (down to $\sim$100~pc-10~kpc from the nucleus) in several nearby radio galaxies
have been investigated.
For example, VLBI images
clarify the detail complex internal structures
such as knot HST-1 in M87 ($z=0.0036$) (e.g., \cite{Cheung2007};\cite{Chang2010};\cite{Giro2012})
and 
knot C80 in 3C~120 ($z=0.033$) (e.g., \cite{Roca2010}; \cite{Agudo2012}),
located 50-100~${\rm pc}$ away from their nucleus (details and other examples are summarized in \S~\ref{classification}).
HST-1 complex, and knots D and E located at kpc order
from the nucleus of M87, display superluminal motions 
up to $6c$ by Very Large Array (VLA), Hubble space telescope (HST) and VLBI
(e.g., \cite{Biretta1995}; \cite{Biretta1999}; \cite{Cheung2007};\cite{Chang2010}; \cite{Giro2012}).
The kinematics of knots in nearby broad line radio galaxy
3C~120 are also studied out to 3~kpc,
but the none of the superluminal motion
originally claimed by \citet{Walker1988}
was found in further VLBI observations
(\cite{Muxlow1991}; \cite{Walker1997}).
As for radio-loud quasars, 
VLBI images at large scales, let alone their kinematics are hardly studied
due to their locations and lack of spatial resolutions.
Although the majority of the cores showing superluminal motions are quasars (e.g., \cite{Kellermann2004}; \cite{Lister2009}),
it is not clear where jet deceleration happens.
To develop a detailed understanding of the jet deceleration process and
shock dissipation process at large scales,
it is crucial to obtain direct images of kpc scale knots with
sufficient spatial resolution.
There is only one attempt to image kpc scale knots and constrain on their motions
in radio-loud quasars.
With a global VLBI network of 16 radio telescopes,
\citet{Davis1991} conducted the observations of 3C~273 ($z=0.158$) at 1.7~GHz.
By comparing with an earlier image,
they indicate a possible superluminal motion
about 2-5$c$ on 100~pc scales,
but it is difficult to confirm because different components emerged.
To overcome the above shown difficulty
and explore observational properties of 
large scale knots in radio-loud quasars, 
we select 3C~380 
which is known as a compact steep spectrum (CSS) radio source
($z=0.692$) with VLBI.
As having a steep spectrum,
this source is considered to be associated with a misaligned jet (\cite{Fanti1990}).
Since the position angle of each inner parsec scale jet, ranging from $284^{\circ}$ to $352^{\circ}$,
are almost parallel to their motion vectors, it is suggested that the jet was ejected ballistically from the core.
There are two distant bright knots, K1 and K2, located $\sim$0.73 and 1 arcseconds at position angle around $308^{\circ}$, which is approximately in the direction of continuation of the inner jet (\cite{Kameno2000}).
The distance between K1 and the core corresponds to more than 20 kpc, using viewing angle 
$\leqq15^{\circ}$ (e.g., \cite{Wilkinson1984}; \cite{Kameno2000}).
Detection of the linear polarization by Multi-Element Radio Linked Interferometer Network (MERLIN) \citep{Flatters1987} and the optical emission by HST \citep{deVries1997} at K1 and K2
implies the presence of strong interaction between the knots and ambient medium.
K1 in 3C~380 is the best target to 
explore the observational properties of quasar kpc scale knot, 
because the knot is sufficiently bright 
and the source entire angular size is sufficiently compact for VLBI observations. 
We attempt to image the kpc scale knot K1 in 3C~380 
with a high resolution VLBI.

The organization of this paper is as follows.
The observation data and data reduction is described in \S~2 and \S~3,
respectively.
The results are presented in \S~4 and
the discussions are given in \S~5.
Throughout this paper, we adopt the following cosmological parameters :
$H_{0}=71~ {\rm km~s^{-1} Mpc^{-1}}$,
$\Omega_{\rm M}=0.27$, and
$\Omega_{\rm \Lambda}=0.73$ \citep{Komatsu2009}, or
$1~{\rm mas}=7.11$~pc and $0.1~{\rm mas~ yr^{-1}}=3.92c$.


\section{Archival Radio Data}
We analyzed two VLBI archival data of quasar 3C~380 at 4.815~GHz in left hand circular polarization, 
observations of which were made on 1998~July~4 and 2001~April~24.
In Table~\ref{data}, we summarize the details of the VLBI data.
All the ten antennas of VLBA (Very Long Baseline Array) are used in the observations.
The first epoch is our data as a part of VSOP (VLBI Space Observatory Programme) observation, including the Effelsberg telescope.
The second epoch data is obtained from VLBA archival data services.
All the correlation processes were performed by 
the National Radio Astronomy Observatory (NRAO) VLBA correlator 
in Socorro, NM, USA.
%


\section{Data Reduction}
We used the Astronomical Image Processing System (AIPS) software package 
developed by the NRAO for a priori amplitude calibration, fringe fitting,
and passband calibration process.
For the first epoch data, we did not use the data of spacecraft baselines.
Since the distance between the core and K1 corresponds to around 700 beamwidth,
time and frequency averaging would cause time and bandwidth smearing in the K1 region (\cite{Thompson2001}).
To minimize the smearing effect on K1 to make wide field of view images, 
the frequency channels were averaged within each IF, 
and these IFs were kept separate during imaging process.
We did not use any time averaging, but
the error on each visibility data point was also adopted as the standard deviation within ten seconds by using AIPS task FIXWT (see Table \ref{data}).

Imaging was performed using CLEAN and self-calibration algorithm. 
This was performed with the Difmap software package (\cite{Shepherd1994}).  
We started imaging the core and the inner jet
to eliminate the sidelobes from the core over K1,
because we expected the K1 flux was around 10\% 
of the integrated flux density of the inner jets, which was $\sim$ 2.0~Jy at 5~GHz.
We adopted both uniform and natural weighting,
and performed phase-only self-calibration several times.
After converging visibility phase model and observed phase, 
we imaged the inner jet and K1 together by applying natural weighting and 
$uv$-tapering and performed phase and amplitude self-calibration.

Since all the data had the shortest $uv$ distance much longer than 143~k$\lambda$,
which corresponded to half of the beam size $\sim$720~mas which covered both core and K1 at 5~GHz,
missing flux would be caused at K1.
Therefore we only obtained the lower limit of the K1 flux.

\section{Results}
\subsection{Inverted Bow-shock Structure of K1}

\begin{figure*}
\begin{center}
\FigureFile(130mm, 80mm){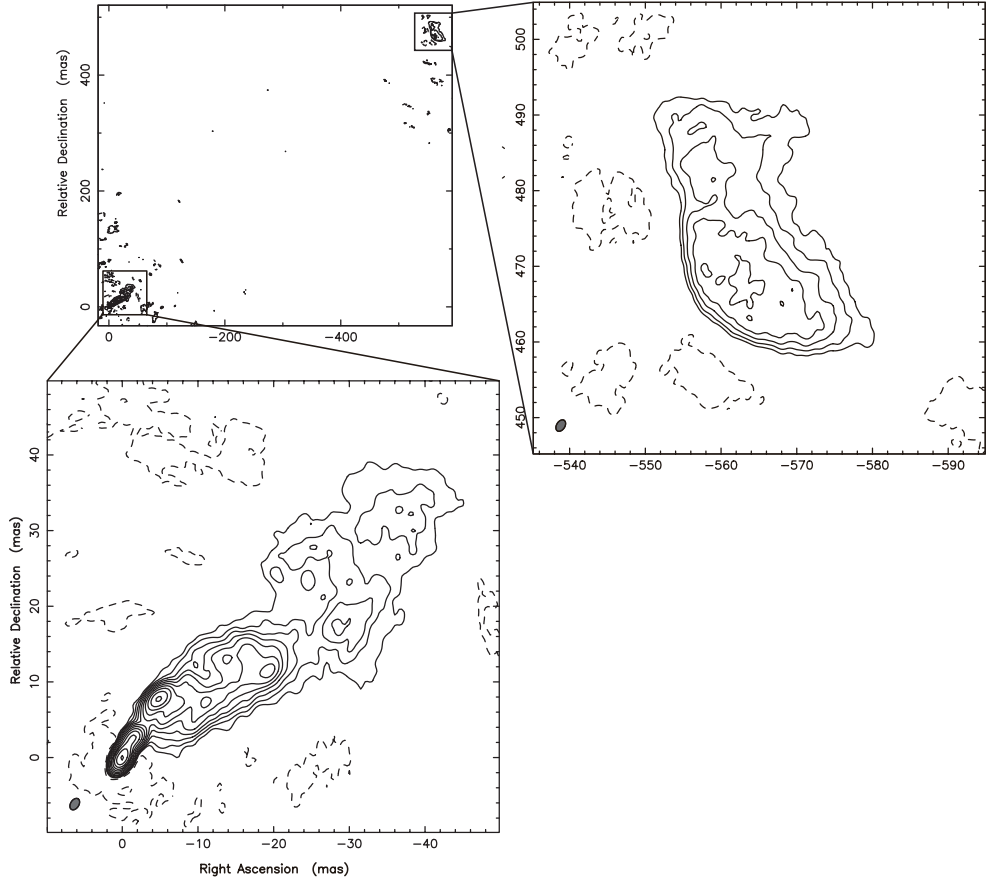}
\caption{Top left panel shows 3C~380 entire image obtained by VLBA ten antennas plus Effelsberg telescope on 1998 July 4 at 4.815 GHz with a resolution of $1.66{\rm~mas} \times 1.10 {\rm~mas}$ in $\rm P.A.=-32.5^{\circ}$, which is shown at bottom left corner of each image. 
The bottom-left panel displays the zoom-in image around the core with contour levels 0.328 $\times$ ($-1$, 1, 1.41, 2, 2.83, 4)  ~$\rm mJy~beam^{-1}$.
The right panel shows the zoom-in image at K1 with contour levels 0.328 $\times$ ($-1$, 1, 2, 4, 8, ..., 2056)  ~$\rm mJy~beam^{-1}$.
Each lowest contour is 3$\sigma$ level.
Natural weighting is applied.}
\label{allmap}
\end{center}
\end{figure*}

In Fig.~\ref{allmap}, we show the overall
images of 3C~380 in total intensities.
In the left panel,
the entire image of 3C~380 and zoom-in core image are shown, 
while the K1 image is shown in the right panel
with the beamsize of 1.66 $\times$ 1.10~mas resolution at beam position angle $-32.5^{\circ}$.
K1 in Fig.~\ref{allmap}, located at around 0.73 arcsec away from the core, 
is detected with signal-to-noise ratio over eight.
From Fig.~\ref{allmap},
we find, with new clarity, 
the inverted bow-shock shaped structure 
in K1 edge-brightened region as is previously described  
(\cite{Simon1990}; \cite{Wilkinson1991}; \cite{Papa2006}).
This finding helps to confirm the original suggestion of 
inverted bow-shock structure in K1 by \citet{Cawthorne2006} and \citet{Papa2006},
with three times better resolution than these of their images.
The width of K1 is about 280~pc (40~mas) 
measured perpendicular to the jet direction, which is the same size as the K1 diameter suggested by \citep{Simon1990},
and the length of K1 is 140~pc (20~mas) in our images.
Compared with the previously obtained VLBI images of K1 (Fig.~12 in \cite{Papa2006}; top left of Fig.~1 in \cite{Kameno2000}),
we have attained the highest spatial resolution image
by adding outer five VLBA and Effelsberg telescopes.
The spatial resolution of obtained K1 image 
is three times higher than that of the previously
obtained highest resolution image of K1 at 1.6~GHz (the beam size of Fig.~12 in \cite{Papa2006} is 5.0 $\times$ 3.7~mas).

\subsection{Kinematics of K1}

\begin{figure*}
\begin{center}
\FigureFile(130mm, 80mm){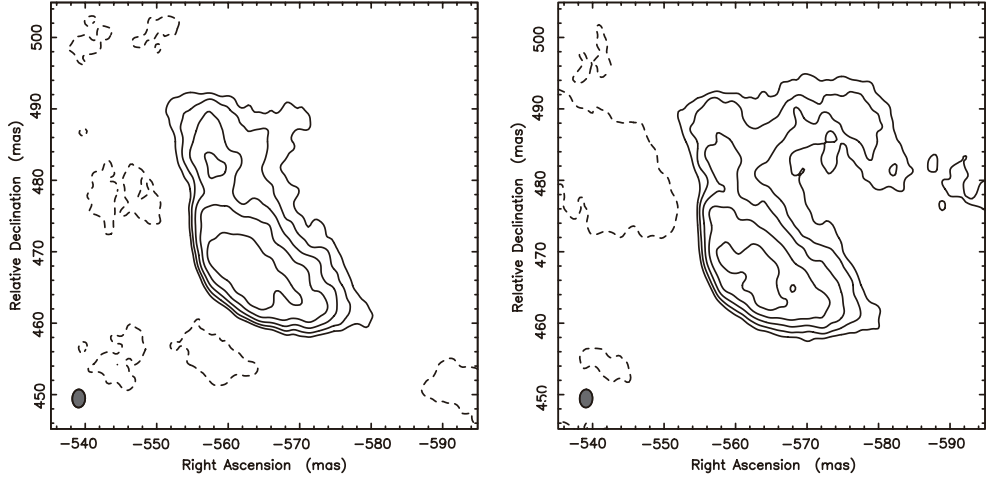}
\caption{Natural weighted images of K1 at 4.815 GHz with
VLBA ten antennas. 
Left image data is obtained on 1998 July 4 and
right image data is on 2001 April 24. 
All beam sizes are restored to $2.54{\rm~mas} \times 1.81{\rm~mas}$ in $\rm P.A. =-1.95^{\circ}$, which is the original resolution of the right map. 
Contour levels are  0.765 $\times$ ($-1$, 1.4142, 2, 2.83, 4) mJy beam$^{-1}$,
which are aligned to the higher 3$\sigma$ level (2001 data).
The details  of these two images are summarized in Tables~\ref{quality} and \ref{K1}.}
\label{k1peakmap}
\end{center}
\end{figure*}

We further attempt to explore the kinematics of K1
by comparing Gaussian-fitted peak positions of K1 in these two epoch images.
In Fig.~\ref{k1peakmap}, 
we present the two epoch images of K1, 
using only VLBA ten antennas, with the common restored 
beam size 2.54 $\times$ 1.81~mas resolution at beam position angle $-1.95^{\circ}$ with natural weighting. 
From Fig.~\ref{k1peakmap}, we find that
the inverted bow-shock structure seen in the first epoch image 
is also detected in the second epoch.
This structure 
would be maintained between these two observations, that is, over 2.82~years.

\begin{figure*}
\begin{minipage}{0.5\hsize}
\begin{center}
\FigureFile(70mm, 40mm){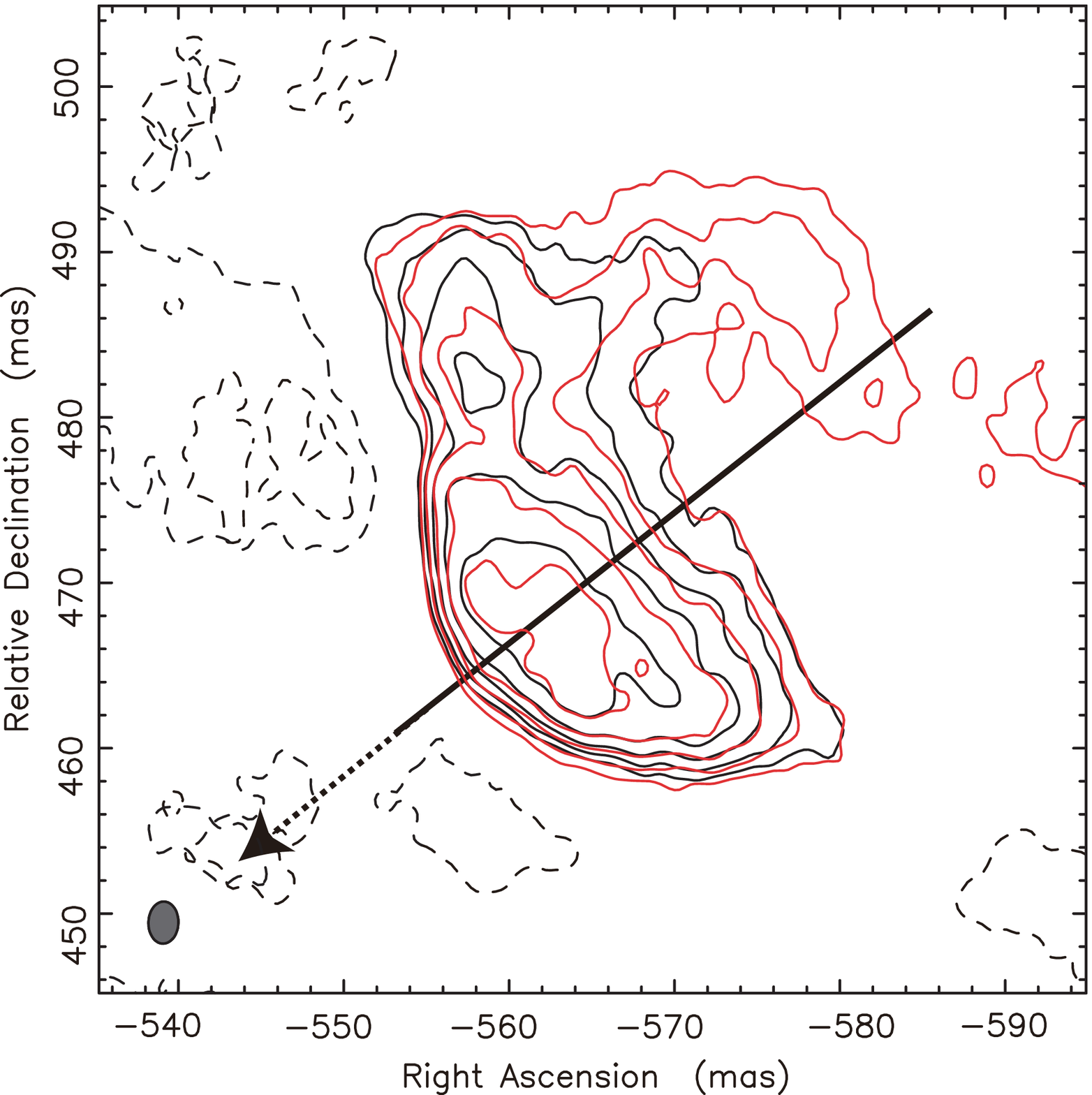}
\end{center}
\end{minipage}
\begin{minipage}{0.5\hsize}
\begin{center}
\FigureFile(80mm, 50mm){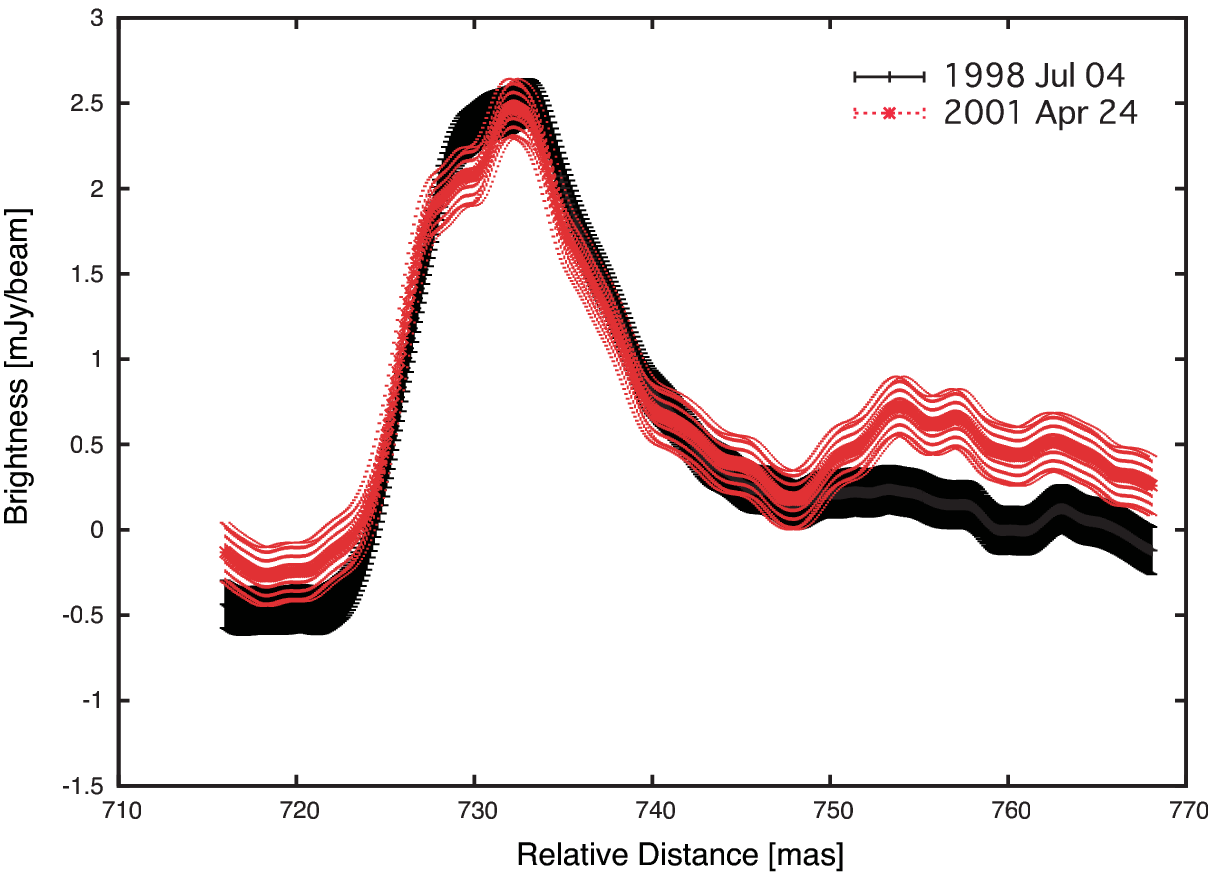}
\end{center}
\end{minipage}
\caption{
Left image is 1998 July 4 image overlaid by 2001 April 24image shown in Fig.~\ref{k1peakmap} with reference to the core brightest peak. 
The straight line shows the slice position.
The slice position is determined along the line connecting the core peak position and mean K1 peak position measured by AIPS task MAXFIT.
The dotted arrow indicates the direction to the nucleus.
Right image shows the slice profiles of K1, along the straight line shown in the left image. 
We put 1$\sigma$ flux error (or image r.m.s. noise in Table~\ref{quality}) on each data point.}
\label{k1overlay}
\end{figure*}

To compare the position of K1 between 1998 July and 2001 April, 
we overlay two images of Fig.~\ref{k1peakmap} to produce Fig.~\ref{k1overlay} left
with reference to the core peak position measured by AIPS task JMFIT.
Previous studies support that 
core brightness peak position converges to a stable point
within 0.2~mas order
(e.g., \cite{OS2009}; \cite{Hada2011}).
Since it is sufficiently small compared with our beam size, 
we can regard the core as stable in the present work.
To measure the peak position and estimate the position accuracy of K1,
we fit a single Gaussian model to each slice profile (Fig.~\ref{k1overlay} right)
by using the task SLFIT in AIPS.
From Fig.~\ref{k1overlay} right, we can find the core-facing edge of K1 in each epoch
is located at the same position, that is, at $\sim$725~mas distance from the core.
The full width of half maximum (FWHM) of 
a single Gaussian fitted to slice profile of K1 is almost identical to the slight change of each image 
due to the method of self-calibration.
The position accuracy of K1 is derived as a ratio of
FWHM of the fitted Gaussian to signal-to-noise ratio (SNR) at K1 (e.g., \cite{Walker1997}),
which is conservatively estimated less than $\sim$0.79~mas.
The derived peak position and their accuracy is summarized in Table \ref{K1}.

Finally we estimate the maximum
apparent velocity of K1, $\beta_{\rm app, max}$, 
as peak position displacement over $\Delta t=$2.82~years with propagation of uncertainty.
The peak position displacement is $\Delta R=R_{1}-R_{2}=-0.27~{\rm mas}$, where $R_{1}$ and $R _{2}$ are the peak position at first epoch and second epoch, respectively.
The error in the displacement is estimated to be $0.97~{\rm mas}$ by using propagation of uncertainty, or the root-mean-square (r.m.s.) of the position uncertainty at each epoch.
Therefore, the upper limit of the peak position displacement is $\Delta R_{\rm max}=-0.27~+~0.97=0.70~{\rm mas}$ and
the maximum apparent proper motion is estimated to be
$\Delta R_{\rm max}/\Delta t=0.25 ~{\rm mas~yr^{-1}}$.
Thus, we obtain $\beta_{\rm app,max}=9.8$.


\section{Discussion}

\subsection{Internal Structures in Large Scale Knots/Hot Spots}

\subsubsection{Classification of previously known cases}\label{classification}

As shown in the Introduction,
VLBI observations of large-scale
knots are quite limited 
and only a handful of sources are explored.
Here, we classify them.
Below, we attempt to categorize the internal structures into 
three typical ones.
We do not include some known sources
which are difficult to categorize because of their peculiarities
(e.g.,  HST-1 in M87 by \cite{Giro2012}; northern hot spot of broad line radio galaxy PKS~$1421-490$ by \cite{Godfrey2009}).

{\it Inverted bow-shock type---}
As shown in the previous section,
the apex of K1 edge-brightened region in 3C~380 faces towards the core.
In this work, we call this feature inverted bow-shock structure.
The same structure as K1 in 3C~380 is found at C80 in 3C~120,
which is the stationary jet feature located 140~pc (80~mas) away 
from the core with 35~pc in size \citep{Agudo2012}.
The key common property between 3C~380 and 3C~120 
is their viewing angle. 
They are classified as misaligned AGN \citep{Abdo2010},
since their viewing angles are larger than those of blazars but
smaller than those of radio galaxies.

{\it Bow-shock type---}
3C~205 is known as a high-redshift quasar
with large viewing angle 
because there exists a pair of strong hot spots.
In the pioneer work of \citet{LB1998},
VLBA images of the primary hot spot 
A in 3C~205 at 1.4~GHz are shown.
The VLBI hot spot, located more than 40 kpc away from the core,
has the overall size 1400~pc
and the jet width around 250~pc.
The apex of edge-bright region in hot spot A of 3C~205 against the core
face the opposite direction of that of 3C~380.
Therefore, here we call this feature  
bow-shock type structure to tell contrast to inverted bow-shock type.
\footnote{Note that
\citet{LB1998} 
focused on the asymmetry of bow-shock
rather than the bow-shock
and they discussed a bent-jet model which 
can explain the asymmetry.}

{\it Multi-spots type---}
There are several knots and spots 
having multi-spots in a hot spot.
Pictor A is a representative of this.
\citet{Tingay2008} reveals that
the northwest hot spot in Pictor A at 3.5~kpc scale contains
five compact pc-scale components in the spot.
The sizes of these components are 
30-170~ pc.
One of the other examples is the
southern hot spot of FRI/FRII radio galaxy PKS~$2153-69$,
which is 200~pc in diameter
and contains three components as small as 50~pc \citet{Young2005}.
The hot spot is located 5~kpc away from the core
and would trace the varying position of the precessing jet interaction region with clouds.

\subsubsection{Origin of various internal structures}

Bearing the above brief summary in mind,
let us discuss possible origins of 
apparently different internal structures in large scale knots.

{\it Viewing angle effect---}
The inverted bow-shock can be observed in
broad-line radio galaxies (BLRGs) and CSS-QSOs, 
both of which have relatively narrow viewing angles.
It is known that 3C~120 is
identified as a BLRG and 
its viewing angle is estimated as 
$\theta\leqq 19^{\circ}$ (e.g., \cite{Gomez2000})
and
CSS-QSO 3C~380 with inclination angle 
$\theta\leqq 15^{\circ}$ (e.g., \cite{Wilkinson1984}; \cite{Kameno2000}).
On the other hand,
the viewing angle of quasar 3C~205 is
suggested to be around 40$^{\circ}$,
which is the upper end of the quasar/radio galaxy unification
according to low lobe flux density ratio \citep{Bridle1994}.
Therefore, we speculate that the difference of
viewing angles divide images into
bow-shock and inverted bow-shock.
This point has been already suggested by \citet{Cawthorne2006},
modeling the edge-bright region in K1 as a conical-shock seen with small viewing angle.
Our work contributes to offer the highest resolution image of K1
structure with new clarity 
and 
to show the inverted bow-shock structure
supporting the Cawthorne's model.

Regarding the physical origin of bow-shock and inverted bow-shock,
\citet{Lind1985} suggest that the bow-shock is caused by a fast stream moving at relativistic speed up the center of the jet,
while for example, \citet{Norman1982} indicate the inverted bow-shock is triggered by  Kelvin-Helmholtz instability inside the unshocked jet.
In the case of K1 in 3C~380,
the inverted bow-shock might be interpreted as the bent backflow (reverse shock) at the jet termination point \citep{Mizuta2010},
since \citep{Wilkinson1991} mention that K1 is similar to a hot spot seen in the lobes of some Fanaroff-Riley class II sources seen approximately pole-on.

{\it Precession effect---}
The jet precession effect, or we may say jet-jittering effect, 
is explored and modeled by \citet{Scheuer1982} and \citet{Cox1991}
and are known as the ``dentist drill" model.
We consider that when the direction of the straight jet changes, 
causing the termination point to vary over a large-scale spot larger than the cross section of the jet,
dynamically young (or long-lived) relic components can be seen as multi-spots.
The multi-spots seen in Pictor A
can be explained by the dynamically
young (or we may say the long-lived) relic
components produced by the precessing jets
\citep{Tingay2008}.
They estimate that a typical
synchrotron cooling time scale of these regions from
100 to 700~years is much longer than
the dynamical (Alfvenic crossing) time scale of 
a few decades and indicate that
these are dynamically young regions. 
%

\subsection{Kinematics of kpc knot K1}
First of all, we stress that 
the present work is the first attempt to constrain the upper limit on
possible proper motion at kpc scales
in radio-loud quasars.
By comparing Gaussian peak position of K1 slice profiles in 1998 July
and 2001 April as reference to the core peak position (\S~4.2), 
we constrain the resolution of K1 apparent proper motion 
up to 0.25$\rm~mas~yr^{-1}$, corresponding to apparent velocity $\beta_{\rm app,max}=9.8$.
In the core region,
proper motions of several components are measured by \citet{Kameno2000} and \citet{Lister2009},
ranging from $1.2c$ to $15c$, from sub-mas to 30~mas away from the core, respectively.
Our constraint is marginally slower than the fastest and outermost apparent motions measured in the core region,
which is the apparent motion of component F, 0.38${\rm~mas ~yr^{-1}}$ or $15c$,
labeled by \citet{Kameno2000}
\footnote{Values are recalculated with the cosmology parameters shown in \S~1}.
This implies the jet deceleration or bending occurs between inner jet and K1, 
or the ejection angle (viewing angle) of K1 has changed from those of the inner jets assuming straight ballistic jets.
To confirm jet proper motion at large scales with the maximum resolution of apparent velocity less than $2c$, further new epoch VLBI observation more than 14 years interval from the first epoch observation is required.
In the case of jet bending, the apparent position angle difference $\phi_{\rm pos}\sim13^{\circ}$ between F and K1 would be magnified by projection with fixed small viewing angle ($\theta_{\rm view}\leqq15^{\circ}$).
Intrinsic jet bending angle $\phi_{\rm bend}$ is estimated to be $\leqq3.3^{\circ}$,
where $\rm tan~\phi_{bend}=tan~\phi_{pos}\times sin~\theta_{view}$ \citep{Kameno2000}.
As for the changes of jet ejection angle, if we assume the same intrinsic velocity $\beta=0.9978$ for F and K1, the viewing angle should be $3.8^{\circ}$ for component F and $1.4^{\circ}$ or $10.2^{\circ}$ for K1.

\subsection{Future prospect}

As a first step, we deal with only 5~GHz VLBI data in this paper.
Here we mention future prospects to investigate the properties of K1.

\subsubsection{Low frequency spectrum turnover}

Low frequency spectrum turnover 
can constrain the jet component 
properties such as magnetic field strength
(e.g., PKS $1421-490$: \cite{Godfrey2009}).
Regarding the case of K1 in 3C~380, 
previous work of \citet{Megn2006}
suggests spectral flattening  
below $\sim100~{\rm MHz}$.
However discussions in \citet{Megn2006}
are based on flux values collected 
from literatures derived from various different interferometers 
and in which K1 is smaller than the beam size of each interferometer.
Therefore, it seems difficult 
to determine fluxes accurately. 
Square Kilometer Array (SKA)
\footnote{http://www.skatelescope.org/}
will, in future, tell us the real turnover frequency
with sufficiently high resolution.

\subsubsection{Polarization properties}

Polarization properties are 
crucial to explore magnetic field geometries.
Only \citet{Papa2006} report
the resolved distribution of 
magnetic vector polarization angle (MVPA) in K1.
The MVPA distribution appears tangential to the
inverted bow-shock.
In order to clarify a change of shock structure in K1, 
time-variation of MVPA is one of the key quantities for future observations
because the sudden change of MVPA strongly 
suggest the existence of magnetohydrodynamical 
fast/slow mode waves (e.g., \cite{Nakamura2011}).
To clarify polarization properties of synchrotron emission
is also substantial \citep{Nalewajko2012} for testing reconfinement shock models (e.g., \cite{Komissarov1997};
\cite{Stawarz2006}; \cite{Bromberg2009}).

\section{Summary}
To explore the properties 
of kpc scale knots in radio-loud quasars,
we produced the pc scale images of
distant knot K1 in a bright CSS quasar 3C~380 with VLBI.
Below we summarize the main results obtained in this work.
\begin{enumerate}

\item
Using VLBA plus Effelsberg telescopes at 5~GHz with the technique of wide field imaging,
we succeed in obtaining the highest resolution images of the pc scale structure of K1,
located at more than 20~kpc downstream of the core. 
We confirm the edge-brightened region in K1 on the side of facing the core as the inverted bow-shock,
which is the clear indication of conical shock
with misaligned viewing angle.
\item
Comparing VLBA ten antennas images of K1 in 1998 July and 2001 April
referencing to the core brightness peak,
the edge-brightened regions are located at $\sim725~\rm mas$.
We constrain the upper limit on the possible proper motion of K1 up to 0.25 $\rm~mas~yr^{-1}$ or $9.8c$.
Since our constraint on the apparent velocity is marginally slower than the fastest knot apparent motions of the core region,
jet deceleration, bending, or precession could have occurred.
Further new epoch VLBI observation
is needed to confirm the proper motion of K1
with the resolution of apparent velocity $\leqq2c$ .

\end{enumerate}

\bigskip
We are grateful to K. Asada and A. Doi for constructive discussions.
We thank the anonymous referee for useful comments and suggestions. 
S.K. acknowledges this research grant provided 
by the Global COE program of University of Tokyo.
This work was partially supported by Grant-in-Aid
for Scientific Researches, KAKENHI 2450240 (MK) from the
Japan Society for the Promotion of Science (JSPS).
This research has made use of data from National 
Radio Astronomy Observatory (NRAO) archive.
The NRAO is a facility of the National Science Foundation operated under cooperative agreement by Associated Universities, Inc.

\clearpage

\begin{table*}[htdp]
\caption{Details of the VLBI observations at 4.815 GHz}
\begin{center}
\begin{tabular}{ccccccccccc}
\hline
Date & code & $t_{\rm on}$\footnotemark[$*$] & Antennas\footnotemark[$\dagger$] &  BW\footnotemark[$\ddagger$] & \multicolumn{2}{c}{Channels}\footnotemark[$\S$]  & $t_{\rm acc}$\footnotemark[$\l$]\\
&  &&&&no. & sep.&  \\
 & & [min] &  & [MHz] &  & [kHz] & [sec] \\
\hline
1998/Jul/04 &V125 & 770 & VLBA10, EB & 16 & 32 & 500 & 4 \\
2001/Apr/24 & W410 & 619 & VLBA10  &16 &128 &125 & 4 \\
\hline
\end{tabular}
\end{center}
{\footnotemark[$*$] Total on source time.}\\
{\footnotemark[$\dagger$] Antennas : VLBA 10=Pie town NM USA, Los Alamos NM USA, Fort Davis TX USA, Owens Vally CA USA, Kitt Peak AZ USA, North Liberty IA USA, Hancock NH USA, Brewster WA USA, Saint Croix VI USA, Mauna Kea HI USA, EB : Effelsberg Germany.}\\
{\footnotemark[$\ddagger$] Total bandwidth.}\\
{\footnotemark[$\S$] Channel numbers per 1 IF and channel width per 1 channel.}\\
{\footnotemark[$\l$] Data accumulation period.}\\
\label{data}
\end{table*}

\begin{table*}[htdp]
\caption{Image performances of Fig. 2 with VLBA ten antennas}
\begin{center}
\begin{tabular}{cccccccc}
\hline
Date &  \multicolumn{3}{c}{Synthesized beam}\footnotemark[$*$]& $S_{\rm tot}$\footnotemark[$\dagger$] & $S_{\rm peak}$\footnotemark[$\ddagger$] & r.m.s.\footnotemark[$\S$] \\
  &	$a_{\rm maj}$& $a_{\rm min}$& P.A.&\\
 &[mas]&[mas]&$[^{\circ}]$&[Jy]& [mJy beam$^{-1}$] &[mJy beam$^{-1}$]\\
\hline
 1998/Jul/04 & 2.37 & 1.97& -12.1&   2.30 & 835 & 0.139\\
 2001/Apr/24 & 2.54 & 1.81& -1.95 & 2.12 & 934 & 0.172\\
 \hline
\end{tabular}
\end{center}
{\footnotemark[$*$] Long axis, short axis, and position angle of synthesized beam.}\\
{\footnotemark[$\dagger$] Total cleaned flux of entire image.}\\
{\footnotemark[$\ddagger$] Peak flux of the entire image.}\\
{\footnotemark[$\S$] Root-mean-square noise of entire map.}
\label{quality}
\end{table*}

\begin{table*}[htdp]
\caption{Properties of K1 in Fig.~\ref{k1peakmap}}
\begin{center}
\begin{tabular}{cccccccccc}
\hline
Date & $S_{\rm K1}$\footnotemark[$*$] & $S_{\rm K1,peak}$\footnotemark[$\dagger$] & ${\rm SNR_{\rm K1}}$\footnotemark[$\ddagger$] & FWHM\footnotemark[$\S$] & Peak position\footnotemark[$\l$]\\
 & [mJy] &[mJy beam$^{-1}$] & & [mas] & [mas]\\
\hline
1998/Jul/04 &  $\geqq$~152 & $\geqq$~2.60 & $\geqq$~18.7 & 10.42 & 732.53~$\pm$~0.56\\
2001/Apr/24 & $\geqq$~163 & $\geqq$~2.41 &  $\geqq$~14.0 & 11.09 & 732.26~$\pm$~0.79\\
\hline
\end{tabular}
\end{center}
{\footnotemark[$*$] Integrated flux of K1 by using AIPS task IMSTAT.}\\
{\footnotemark[$\dagger$] Peak flux of K1 derived from the Gaussian model fitted to each slice profile by using AIPS task SLICE and SLFIT.}\\
{\footnotemark[$\ddagger$] SNR of K1, or peak flux divided by r.m.s. noise in Table~\ref{quality}. The residual after fitting Gaussian to each slice profile is as small as the r.m.s. noise of the entire map.}\\
{\footnotemark[$\S$] FWHM of Gaussian model fitted to K1 slice profile using SLFIT.}\\
{\footnotemark[$\l$] K1 peak position measured by SLFIT.
The reference position is core brightness peak. 
The position error is FWHM over SNR of K1.}
\label{K1}
\end{table*}

\clearpage

\clearpage

\clearpage

\end{document}